\documentclass[12pt,a4paper]{article}

\usepackage[utf8]{inputenc}
\usepackage{makeidx}
\usepackage{graphicx}
\usepackage{amsmath}
\usepackage{subcaption}
\usepackage{hyperref}

\author{Amir Arslan Haghrah, Amir Aslan Haghrah}

\title{Scalability of Morality: A Particle-Based Numerical Study on the Decoupling of Law and Ethics in Large-Scale Populations}

\begin{document}

\maketitle

\begin{abstract}
This study introduces a particle-based computational framework to investigate the scalability of morality and the systemic decoupling of formal law from decentralized social ethics in expanding populations. While micro-societies reinforce ethical conduct through local reciprocity, macroscale systems introduce anonymity that strains cognitive memory limitations. We model individual agents as discrete particles with finite memory capacities ($L$) and dynamically evolving, stochastic choice profiles ($\mu$) regulated by non-linear social pressure switches. Monte Carlo ensemble simulations demonstrate a distinct, non-linear phase transition as the population scales ($N \to \infty$). When the population metric outpaces memory capacity ($N \gg L$), the local re-encounter probability drops as $\mathcal{O}(L/N)$. This structural dilution neutralizes decentralized peer-to-peer accountability, causing global behavioral norms to decouple from moral baselines and drift toward a minimalist legal floor. Furthermore, cyclic scale experiments expose a prominent, path-dependent hysteresis loop, mathematically formalizing the non-Markovian inertia and irreversible nature of moral decay in self-organizing social systems.
\end{abstract}

\section{Introduction}
The dual nature of social order is anchored in the distinction between legality and morality \cite{kant_metaphysics_2017}. While ethics represents the internal normative framework that guides individual conduct toward collective well-being, legal systems serve as the external, codified apparatus designed to enforce social cooperation through a structured regime of rewards and punishments \cite{hart1961concept,bentham1789introduction}. Historically, the stability of the human civilisation has relied on the alignment of these two pillars. However, a critical divergence exists: legal frameworks are often "minimalist" by design, ensuring only the prevention of gross violations while leaving a vast territory of "unethical but legal" behaviors open for exploitation \cite{stone1975where, carroll1991pyramid}. In such spaces, individuals may derive significant utility from actions that, while technically permissible under the law, fundamentally erode the ethical fabric of the community \cite{pistor2019code}.

The efficacy of moral regulation is inherently sensitive to the architecture of the society in which it operates. In small-scale, high density social configurations, ethical behavior is self-reinforcing through a decentralized system of individual sanctions and rewards \cite{ostrom1990governing}. In these "micro-societies," the high probability of recurrent interactions creates a robust feedback loop of accountability \cite{axelrod1981evolution}; the social cost of an unethical act, manifested through reputation loss or direct retaliation, often outweights the immediate gain \cite{nowak2006five}. However, as societies scale toward massive, complex systems, a fundamental phase transition occurs \cite{dunbar1992neocortex}.

This study addresses the "Scalability of Morality" by examining how population growth decouples formal law from ethical behavior. As the number of interacting particles increases, the anonymity afforded by system scale significantly dilutes individual-level sanctions \cite{traulsen2009exploration}. In large-scale systems, stochastic interactions ensure an unethical actor is unlikely to re-encounter the same victim, neutralizing individual retribution networks \cite{Perc2017statistical}. Consequently, the burden of maintaining social integrity shifts entirely to the legal framework, which is structurally ill-equipped to govern nuanced ethical conduct. Utilizing a numerical framework grounded in particle dynamics and Monte Carlo simulations, we quantify this decoupling as a distinct nonequilibrium phase transition \cite{Sarker2025alignment}. Identifying the mathematical threshold at which moral accountability collapses is a vital engineering necessity for designing resilient governance structures in an era of unprecedented global connectivity and urban density.

Following this conceptual framework, the remainder of this introduction provides a concise review of existing literature to contextualize our contribution, followed by a summary of the paper's novelty and structural organization. The quantitative study of the social behavior has its roots in Sociophysics, where the interactions of individuals are modeled using the principles of statistical mechanics \cite{jusup2022social}. Early models, such as the \textit{Ising model} or \textit{Schelling's segregation model}, demonstrated that macroscopic social patterns emerge from microscopic rules of interaction \cite{castellano2009statistical}. Recent advancements in Computational Social Science \cite{lazer2020computational} have expanded these concepts to study the evolution of cooperation under varying structural and systemic complexities \cite{han2025cooperation}.

Building upon this paradigm, recent literature in Computational Social Science has shifted toward decoding how systemic architecture and communication limits modify cooperative equilibria. Large-scale global data sets and advanced data processing tools have allowed researchers to document a profound tension between growing institutional complexity and localized social cohesion \cite{Conte2012manifesto}. While macro-level structures frequently rely on generalized institutional rules to sustain large populations, behavioral experiments integrated with agent-based models reveal that decentralized monitoring functions efficiently only when individual interactions remain visible and localized \cite{bentham1789introduction}. When these interactions are stretched across expansive, sparse digital or physical networks, the cognitive and structural limits of informal social control become a primary bottleneck for collective stability \cite{Eynon2025computational}.

To precisely isolate these dynamics, modern studies frequently turn to specialized multi-agent modeling frameworks. In these environments, autonomous decision-making entities are governed by heterogeneous behavioral rule specifications, allowing for the precise analysis of emergent phenomena across varying spatio-temporal scales \cite{Jamali2026agent}. Recent simulations mapping the evolution of pre-state communities indicate that while decentralized information exchange and gossip-driven reputation assessment can suppress opportunism in small configurations, long-term systemic scalability invariably triggers a structural reliance on centralized, coercive frameworks \cite{Vasellini2024applications}. Consequently, computational modelers have increasingly positioned agent-based frameworks as vital simulation sandboxes to test the bounds of institutional policy and public governance prior to structural deployment \cite{Belfrage2024simulating}.

This collapse of localized moral accountability under rapid scaling can be formally mapped to the physics of phase transitions. Within nonequilibrium statistical mechanics, collective human behavior is modeled as an active matter system where microscopic adjustments in density or proximity induce spontaneous macroscopic symmetry breaking \cite{Sarker2025alignment}. Simulating spatial evolutionary games via Monte Carlo updates reveals that shifting from localized lattices to massive, well-mixed networks yields sharp bifurcations between universal cooperation and systemic defection \cite{Menon2018emergence}. Furthermore, multi-population mean-field theories confirm that when agent density thresholds cross a critical value, the global order parameter experiences an abrupt jump \cite{Contucci2008phase}. Our paper contributes directly to this intersection by using particle dynamics to mathematically bound the operational threshold at which decentralized moral regulation structurally fails.

Previous research in the field of Sociophysics has successfully used Monte Carlo methods to model phenomena like opinion formation and cultural dissemination. These studies demonstrate that the "global state" of a system is highly sensitive to the local "interaction density" \cite{jusup2022social}. Furthermore, the literature on Agent-Based Modeling (ABM) has explored how local feedback loops, such as individual rewards and punishments, maintain system cooperative stability \cite{gupta2025role}. However, a significant gap remains: most models assume a uniform enforcement of rules, failing to account for the divergent dynamics between centeralized legal constraints and stochastic individual interactions \cite{Koid2025agent}.

While the "Anonymity Effect" in large-scale systems has been qualitatively discussed in social sciences, it lacks a rigorous numerical characterization regarding how group size actively alters individual behavioral strategic payoffs \cite{Pereda2019group}. Specifically, how the scaling of a system ($N \longrightarrow \infty$) physically "dilutes" the impact of local individual sanctions, leading to a decoupling from global legal frameworks, is a problem of systemic stability that has yet to be fully quantified through the lens of particle-based simulations and nonequilibrium spin models \cite{Perc2017statistical,Baronchelli2018emergence}.

The evolution of complex Cyber-Physical-Social Systems (CPSS) underscores the difficulty of modeling human behavior under deep systemic uncertainty \cite{wang2022new}. Because data-centric learning models often struggle to integrate domain expertise with behavioral ambiguities, contemporary literature emphasizes specialized mathematical frameworks—such as fuzzy logic and decentralized architectures—to bridge the gap between microscopic agent behaviors and macroscopic social trends \cite{wang2022new}. Moving away from black-box predictive models, this study introduces a particle-based stochastic framework that maps non-linear social pressures directly onto state-space dynamics.

Our approach builds upon the established paradigm of modeling competitive socio-behavioral dynamics as non-linear, parameter-driven systems rather than relying on empirical big data. As demonstrated by Li et al. \cite{li2022modeling}, formalizing social phenomena like opinion propagation as evolving network parameters offers deep theoretical insights into stability, convergence, and attractor dominance. Extending this methodology to normative dynamics, we treat individual ethical propensities ($\mu$) and motivational velocities ($\rho$) as coupled dynamic parameters to evaluate how macroscale structural changes drive a self-organizing population toward competing, path-dependent equilibria.

This article departs from qualitative sociological narratives and static optimization models, offering a dynamic, numerical prespective on social ethics. The primary innovations are:
\begin{itemize}
    \item \textbf{Formalization of Decoupling Dynamics:} We define a mathematical framework to observe how the correlation between ethical behavior and legal compliance weakens as a function of system entropy and population scale.
    \item \textbf{Stochastic Interaction Modeling:} Using a particle-based approach, we simulate the probability of encounter-based sanctions, providing a more realistic representation of social accountability than traditional mean-field approximations.
    \item \textbf{Identification of the Critical Scaling Point:} We identify the numerical threshold where the decentralized "ethical feedback" fails, leaving the system's integrity solely dependent on centralized (legal) control.
\end{itemize}

The remainder of this paper is organized as follows: In Section \ref{sec:architecture}, the system architecture and modeling are presented, detailing the mathematical formulation of the particle system, the transition rules for rewards and punishments, and the Monte Carlo setup. In Section \ref{sec:simulations}, we provide the simulation results and sensitivity analysis, which analyze the behavior of the model under varying population densities and examine the decoupling phenomenon. Section \ref{sec:discussion} offers a systemic discussion, interpreting the findings within the context of control theory and social stability. Finally, Section \ref{sec:conclusion} summarizes the results and outlines future directions for large-scale social modeling.

\section{System Architecture and Modeling}
\label{sec:architecture}
The proposed model simulates a society as a complex system of interacting particles, where each particle represents an individual agent capable of making stochastic decisions regarding ethical conduct. The architecture is designed to capture the interplay between individual memory, personal motivation, and the overarching social environment.

\subsection{Agent Architecture and the Ethical State Space}
Each agent in the system, defined as a Particle, is characterized by an internal state vector known as the Ethical View. Unlike static models of cooperation, the Ethical View is a dynamic construct that evolves based on the agent's history of interactions. The core of this state is defined by the parameter $\mu[k] \in \left[0, 1\right]$, representing the Unethical Behavior Probability in $k^{th}$ iteration. At any given interaction in $k^{th}$ iteration, an agent chooses an unethical path with probability $\mu[k]$ and an ethical path with probability $1-\mu[k]$.

The evolution of $\mu[k]$ is not arbitrary but is governed by a secondary internal variable, $\rho[k]$, which represents the Cumulative Motivational Rate of Change. Mathematically, the tendency of an agent to shift toward or away from ethical behavior is dictated by the magnitude and sign of $\rho[k]$. To ensure that the probability $\mu[k]$ remains bounded within the logical $\left[0, 1\right]$ interval, we employ a logistic-style growth equation for its update:
\begin{equation}
    \mu[k+1] =\mu[k] + \Delta \rho[k] \mu[k] \left( 1 - \mu[k] \right)
\end{equation}
where $\Delta$ is the simulation time step. This formulation ensures that the ethical state demonstrates "saturation" effects near the boundaries, representing the increased resistance to change when an agent is either deeply entrenched in ethical conduct or fully committed to unethical behavior.

\subsection{Memory Dynamics: The L-Length Finite Horizon}
A critical feature of the system architecture is the inclusion of a finite memory for each agent, implemented through a Whitelist and a Blacklist. Each particle maintains these lists with a maximum length L, representing the cognitive or social limits of an individual's memory regarding past encounters.
\begin{itemize}
    \item \textbf{Whitelist}: Contains identifiers of agents who have previously acted ethically toward the particle.
    \item \textbf{Blacklist}: Contains identifiers of agents who have committed unethical acts toward the particle.
\end{itemize}
As a particle system, this memory structure introduces a "Sallen-Key" style filtering effect on social interactions: as the population $N$ grows significantly larger than the memory capacity $L$, the probability of an agent re-encountering a specific individual from their memory diminishes. This structural constraint is the primary mechanism through which population scaling decouples individual ethical sanctions from global social behavior.

\subsection{Interaction Logic and Stochastic Feedback Loops}
The societal dynamics are driven by pairwise interactions between particles, selected through a random shuffling mechanism to simulate the stochastic nature of social encounters. During each interaction between a primary agent ($p_{1}$) and a secondary agent ($p_{2}$), the behavior of $p_{1}$ is determined stochastically by its current unethical probability $\mu[k]$. The system distinguishes between two primary categories of social encounters, each governed by specific feedback functions that map the event into a change in the motivational rate, accumulating directly into $\rho$:

\begin{enumerate}
    \item \textbf{Initial Interactions:} If $p_{1}$ is not present in $p_{2}$'s memory records (residing in neither the whitelist nor the blacklist), the interaction is treated as an unestablished social encounter. The system handles this regime through the following flattened sequence:
    \begin{itemize}
        \item \textbf{Ethical Action:} If $p_1$ acts ethically, $p_2$ appends $p_1$ to its localized \textbf{Whitelist}.
        \item \textbf{Unethical Action:} If $p_1$ acts unethically, $p_2$ appends $p_1$ to its localized \textbf{Blacklist}.
        \item \textbf{Rate Accumulation:} The motivational feedback rates for both participating particles are updated concurrently via the functions $\eta$ (for the primary actor) and $\gamma$ (for the secondary observer or victim).
    \end{itemize}

    \item \textbf{Re-interactions (Reciprocity):} If $p_1$ is already structurally recognized by $p_2$, the encounter bypasses initial learning and triggers a \textbf{Reciprocal Feedback} mechanism governed by the function $\lambda$. The direction, sign, and magnitude of this feedback are modeled using a single level of structural possibilities:
    \begin{itemize}
        \item \textbf{Altruistic Gifting ($\lambda_{\text{pos}} < 0$):} Occurs during a positive re-interaction when $p_1$ acts ethically while present on $p_2$'s whitelist within a cooperative society. Ethical behavior is actively rewarded or "gifted" by peers, driving down the motivational rate $\rho$ and decreasing future unethical choices.
        \item \textbf{Ethical Deprivation ($\lambda_{\text{pos}} > 0$):} Occurs during a positive re-interaction under predatory socio-economic conditions. Here, an agent experiences an ethical deprivation cost, missing out on the material gains of exploitation while witnessing unpunished unethical actors being rewarded, paradoxically accelerating the agent's transition toward opportunism.
        \item \textbf{Neutral Indifference ($\lambda_{\text{pos}} = 0$):} Occurs during a positive re-interaction where routine ethical conduct yields no motivational modification, representing an environment where compliance is taken for granted without generating structural social capital or friction.
        \item \textbf{Negative Punishment Dynamics ($\lambda_{\text{neg}}$):} Occurs during a negative re-interaction when $p_1$ acts unethically and is already recognized on $p_2$'s blacklist. This triggers a strong, stabilizing negative reinforcement value designed to simulate localized social punishments, ostracization, or peer-enforced reputational sanctions.
    \end{itemize}
\end{enumerate}

\subsection{Mapping Functions and Social Pressure}
The internal update functions ($\eta$, $\gamma$, $\lambda$) serve as the core control parameters of the dynamical system. To model the non-linear influence of the collective macroscopic environment on microscopic individual motivation, we utilize a \textbf{Hyperbolic Tangent-based Switch Function} for $\eta$ and $\gamma$:
\begin{equation}
    \label{mapping_function}
    f\left(x; c, s, a, b\right) = \frac{(b - a)}{2} \left[ 1 - \tanh\left(s \left(\frac{x}{c} - 1\right)\right) \right] + a
\end{equation}
In this formulation, $x$ represents the \textbf{Mean Unethical Probability} ($\mu_{mean}$) of the society, $c$ defines the critical threshold center, $s$ dictates the scaling slope of the transition, while $a$ and $b$ represent the lower and upper asymptotic bounds, respectively. 

This formulation acts as a non-linear regulatory mechanism that tracks environmental feedback:
\begin{itemize}
    \item \textbf{Sub-threshold Regime (micro-scale or clean societies):} When $\mu_{mean} < c$, the function approaches its upper asymptotic bound $b$. For an agent committing an unethical act, this yields a positive motivational return accumulating on $\rho$, indicating that low systemic corruption provides an open, unpoliced window where individual temptation can freely accelerate.
    \item \textbf{Super-threshold Regime (macro-scale or corrupt societies):} As the society passes the critical threshold $c$, the widespread deterioration of social norms activates environmental resistance or formal systemic pressures. The function drops toward its lower asymptotic bound $a$ ($a < 0$), introducing a negative feedback force on $\rho$, that suppresses further unconstrained growth of individual unethical probabilities.
\end{itemize}

\subsection{Damping and Decay of Motivation}
To prevent the system from experiencing infinite motivational growth and to simulate the "cooling off" period after a social interaction, we introduce a \textbf{Damping Factor} $\phi$. Following each iteration of the society, every agent's motivational rate $\rho$ is updated as:
\begin{equation}
    \rho[k+1] = \phi \rho[k]
\end{equation}
where $0 < \phi < 1$. This ensures that while an interaction can cause a sharp spike in an agent's tendency to change their behavior, this impulse decays over time unless reinforced by further social contact.

\subsection{Global Iteration and Social Shuffling}
The temporal evolution of the society is modeled as a series of discrete time-steps. In each iteration, a "Social Shuffling" process occurs to ensure the stochasticity of encounters. The population $N$ is randomly permuted, and a subset of particles, defined by the number of interactions in each iteration, is paired for collision. This mechanism ensures that the interaction density is controlled and that no agent is artificially locked into a fixed neighborhood, mirroring the fluid nature of mass societies.
Following the pairing phase, the system performs a global update of all internal states. The calculation of the Global Ethical Metric ($\mu_{mean}$) is performed at the end of each step:
\begin{equation}
    \mu_{mean}[k] = \frac{1}{N} \sum_{i=1}^{N} \mu_{i}[k]
\end{equation}
This value serves as the primary state variable of the society, acting as the global feedback signal that influences individual motivational functions in subsequent steps.

\subsection{Monte Carlo Framework and Statistical Convergence}
To account for the high sensitivity of the system to initial conditions and the stochastic nature of the hyperbolic tangent switch functions, a Monte Carlo (MC) methodology is employed. A single execution of the code (running for $20,000$ iterations) represents one realization of the social trajectory.
To achieve statistical significance and minimize the impact of "stochastic outliers," the simulation is repeated across a large ensemble of independent runs. Each realization begins with:
\begin{itemize}
    \item \textbf{Heterogeneous Initialization:} Initial ethical probabilities ($\mu$) and damping rates ($\phi$) are assigned using a random distribution, ensuring the population is not a monolith but a collection of diverse ethical profiles.
    \item \textbf{Variable Memory Allocation:} The memory length $L$ for each particle is slightly randomized to reflect varying cognitive capacities for social record-keeping.
\end{itemize}
The results from these multiple runs are averaged to compute the Expected Social Behavior. This ensemble averaging allows us to identify the "Steady State" of the society and observe the Phase Transition from ethical stability to legal decoupling. By sweeping the population parameter N across different Monte Carlo sets, we can precisely quantify the point at which the society's memory-based ethical enforcement collapses, leaving the global ethical state to drift toward the "unethical but legal" attractor.

\subsection{Sensitivity and Stability Analysis}
Finally, the Monte Carlo setup is used to perform a sensitivity analysis on the mapping functions $\eta$, $\gamma$, and $\lambda$. By adjusting the slopes ($s$) and centers ($c$) of the switch functions, we observe how resilient the society is to "ethical shocks." This numerical approach allows us to treat ethics not as a philosophical absolute, but as a self-organizing system whose stability is a function of interaction frequency, individual memory, and population scale.

\section{Simulations and Analysis}
\label{sec:simulations}

To validate the theoretical framework of moral scalability and the decoupling of law from ethical behavior, we executed a series of numerical experiments utilizing the detailed multi-particle stochastic simulation platform. The simulation parameters, including the specific mappings for the switch functions ($\eta, \gamma, \lambda$), the damping rate $\phi$, and initialization thresholds, were tuned to explore the phase space of the system. 

Specifically, the baseline multi-agent ensemble is structured around a heterogeneous population of $N$ particles, partitioned into two distinct sub-populations to capture structural socio-behavioral diversity:
\begin{equation}
    P_i \in 
    \begin{cases} 
        \text{Group 1 } (N_1 = \frac{7}{8} N), & \text{governed by } \eta_1, \gamma_1, \lambda_1 \\ 
        \text{Group 2 } (N_2 = \frac{1}{8} N),  & \text{governed by } \eta_2, \gamma_2, \lambda_2 
    \end{cases}
\end{equation}
In which, $i$ is the individual index. The continuous temporal domain is discretized using a uniform simulation time step of $\Delta = 0.1$, with each unique realization spanning a relaxation horizon of $20,000$ global iterations to guarantee asymptotic convergence into a statistical steady state. Individual cognitive storage limits are governed by a base memory length parameter of $L$, subjected to a randomized scaling allocation modeled as:
\begin{equation}
    L_i = \lfloor \left( 0.75 + 0.25 \mathcal{U} \right)L \rfloor
\end{equation}
where $\mathcal{U}$ represents a random number in $\left[0,1\right]$ interval with uniform continuous distribution used to induce structural heterogeneity across individual memory horizons. Particle properties are initialized dynamically using stochastic profiles bounded by maximum operational thresholds for the initial state attributes:
\begin{equation}
    \begin{cases}
        \mu_i[0] \sim \mu_{\text{max}}\mathcal{U}, & \mu_{\text{max}} = 0.3 \\
        \phi_i \sim \phi_{\text{max}}\mathcal{U},     & \phi_{\text{max}} = 0.8
    \end{cases}
\end{equation}
while setting the initial motivational acceleration rate to a uniform rest state of $\rho_i[0] = 0$ for all particles. During each discrete temporal step, a collision density of $\lfloor N / 2 \rfloor$ random pairings is executed via a uniform shuffling index sequence, establishing a high-fidelity stochastic network landscape that prevents localized spatial locking.

The core microscopic behavioral updates are dictated by the environmental switch parameters evaluated at the end of each iteration loop. For Group 1, representing the standard sociological segment, the mapping profiles are defined as:
\begin{equation}
    \eta_1(\mathcal{I}) = 
    \begin{cases} 
        0.0, & \mathcal{I} = ethical \\ 
        f(\mu_{\text{mean}}; 0.5, 2.0, -0.2, 0.2), & \mathcal{I} = unethical 
    \end{cases}
\end{equation}
\begin{equation}
    \gamma_1(\mathcal{I}) = 
    \begin{cases} 
        0.0, & \mathcal{I} = ethical \\ 
        f(\mu_{\text{mean}}; 0.5, 2.0, -0.2, 0.2), & \mathcal{I} = unethical 
    \end{cases}
\end{equation}
\begin{equation}
    \lambda_1(\mathcal{I}) = 
    \begin{cases} 
        0.0, & \mathcal{I} = ethical \\ 
        -0.6, & \mathcal{I} = unethical
    \end{cases}
\end{equation}
In which, $\mathcal{I} \in \left\lbrace ethical,\ unethical \right\rbrace$ is the interaction type and $f(x; c, s, a, b)$ represents the hyperbolic tangent switch function defined in equation (\ref{mapping_function}). Conversely, Group 2 models a highly specialized behavioral segment characterized by asymmetric sensitivity and tighter behavioral bounds, operating under the following functional mapping criteria:
\begin{equation}
    \eta_2(\mathcal{I}) = 
    \begin{cases} 
        0.0, & \mathcal{I} = ethical \\ 
        f(\mu_{\text{mean}}; 0.5, 2.0, -0.1, 0.0), & \mathcal{I} = unethical 
    \end{cases}
\end{equation}
\begin{equation}
    \gamma_2(\mathcal{I}) = 
    \begin{cases} 
        -0.025, & \mathcal{I} = ethical \\ 
        f(\mu_{\text{mean}}; 0.5, 2.0, -0.05, 0.05), & \mathcal{I} = unethical
    \end{cases}
\end{equation}
\begin{equation}
    \lambda_2(\mathcal{I}) = 
    \begin{cases} 
        -0.1, & \mathcal{I} = ethical \\ 
        -0.6, & \mathcal{I} = unethical
    \end{cases}
\end{equation}

\subsection{Parametric Sensitivity and Systemic Bifurcation Analysis}

To systematically evaluate the structural resilience of the proposed multi-particle framework against severe environmental perturbations, we subjected the mixed population to a coordinated, non-localized behavioral disruption, an "ethical shock." Right after initialization, a randomized fraction ($\approx 50\%$) of the aggregate particle ensemble had their individual unethical probability dynamically inflated via $\mu_i \leftarrow \min(0.95, \mu_i + 0.40)$, artificially simulating a sudden collapse of legal deterrence or an injection of systemic corruption. The capacity of the decentralized moral ecosystem to absorb, damp, or catastrophically cascade this structural trauma was explored across five independent Monte Carlo parameter regimes sweeping the center ($c$) and slope ($s$) dimensions of the non-linear hyperbolic tangent regulator $f(x; c, s, a, b)$ defined in equation (2).

\begin{figure}[t!]
    \centering
    \includegraphics[width=0.8\linewidth]{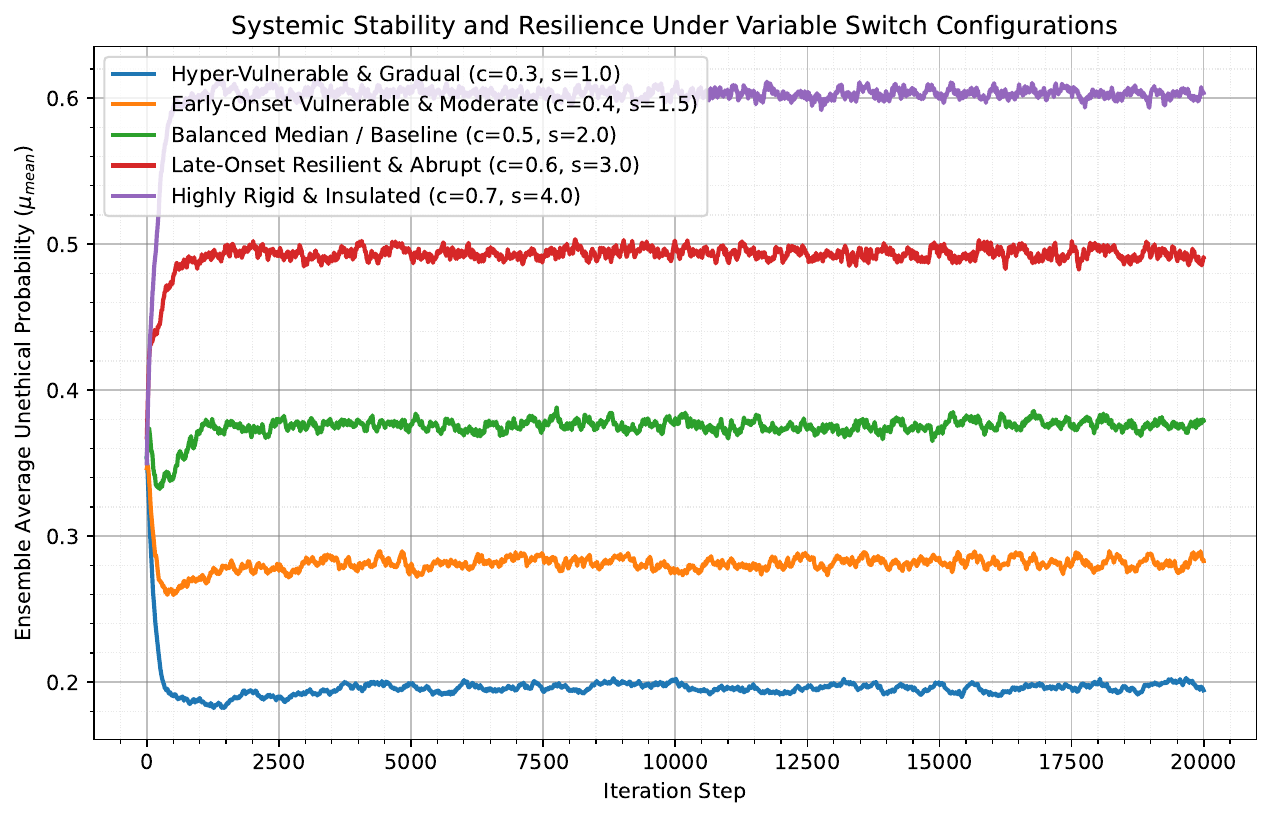}
    \caption{Ensemble average trajectory profiles.}
    \label{fig:sensitivity_analysis}
\end{figure}

The ensemble trajectory averages over independent stochastic realizations reveal a stark, structurally defined bifurcation of global steady states ($\mu_{\text{mean}}$), illustrated comprehensively in Figure \ref{fig:sensitivity_analysis}.

The emergent asymptotic behaviors can be classified into three distinct structural regimes dictated by the systemic sensitivity thresholds:

\begin{enumerate}
    \item \textbf{Hyper-Sensitive Restorative Regimes ($c=0.3, s=1.0$ and $c=0.4, s=1.5$):} Under these configurations, the society features an early-onset regulatory trigger. Despite the substantial initial shock magnitude, the global state vector rapidly relaxes downward within the first $T \approx 1,200$ steps, converging cleanly onto an ethically optimized steady state ($\mu_{\text{mean}} \approx 0.20$ and $\mu_{\text{mean}} \approx 0.28$, respectively). Because the center threshold $c$ is low, minor levels of collective deviation are sufficient to trigger strong negative shifts in the motivational acceleration rate ($\rho \ll 0$) through the microscopic functions $\eta$ and $\gamma$. Individual opportunistic behavior is fiercely penalized during early encounters, driving the network toward complete macro-stabilization before temporary interactions can harden into unpunished habits.
    
    \item \textbf{Balanced Baseline Regime ($c=0.5, s=2.0$):} This parameterization acts as a symmetrical median. The socio-behavioral system displays standard tipping-point characteristics, absorbing part of the shock but showing bounded structural degradation. The ensemble curve settles stably around an intermediate equilibrium of $\mu_{\text{mean}} \approx 0.38$. The system remains homeostatic, but its capacity to fully restore the pre-shock ethical equilibrium is partially degraded due to the sharper transition boundary.
    
    \item \textbf{Rigid Insulated Cascading Regimes ($c=0.6, s=3.0$ and $c=0.7, s=4.0$):} In direct contrast to the sensitive variants, these settings trigger an irreversible behavioral cascade. The system trajectories spike abruptly to high-entropy, low-trust attractors, stabilizing permanently at $\mu_{\text{mean}} \approx 0.49$ and $\mu_{\text{mean}} \approx 0.60$. Because the collective switch center $c$ is set high and the slope $s$ is steep, the peer-enforced moral regulation mechanism remains entirely dormant and "insulated" during the early stages of widespread decay. Particles freely exploit the absence of local social sanctions. By the time the escalating collective metric crosses the high threshold boundary, opportunistic conduct has already normalized into a dominant, absorbing strategy across the network, locking the society into a permanent legal-moral decoupling phase.
\end{enumerate}

Crucially, the narrow standard error bounds and the rapid transition to a bounded, oscillatory statistical steady state for all trajectories after approximately 2,500 iterations demonstrate that long-term societal configuration stability is an explicit function of early-stage feedback sensitivity rather than late-stage rigid resilience.

\subsection{Phase Transition and The Decoupling Threshold}
The first simulation scenario systematically swept the population size $N$ across a geometrically spaced range from $10^1$ to $10^3$ to observe the macro-societal steady state under a fixed base memory capacity $L = 20$. For each population scale, a Monte Carlo ensemble of $20$ independent realizations was executed for $20,000$ iterations to ensure asymptotic convergence.

\begin{figure}[t]
  \centering
  \includegraphics[width=0.8\linewidth]{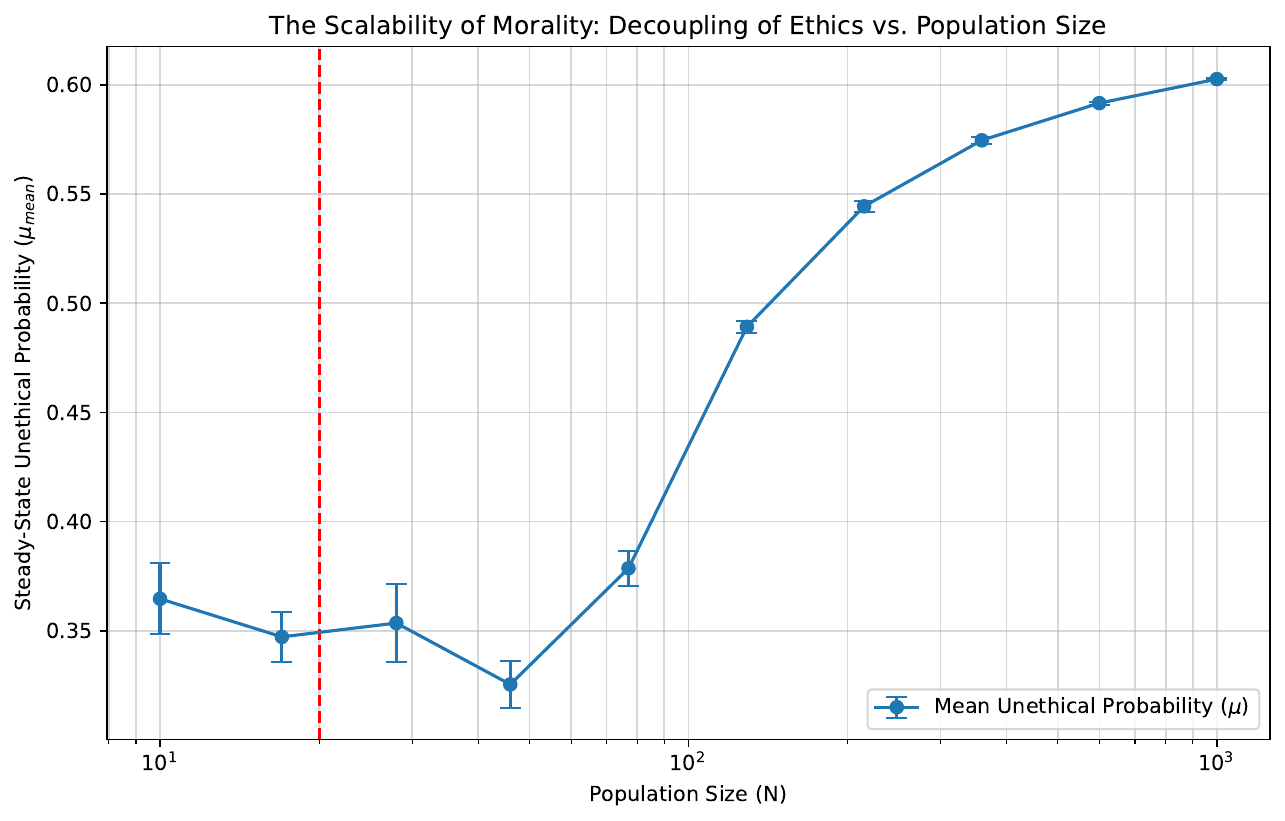}
  \caption{Phase transition and the decoupling threshold diagram.}
  \label{fig:phase_transition_diagram}
\end{figure}

As illustrated in the phase diagram, Figure \ref{fig:phase_transition_diagram}, the system exhibits distinct behavioral regimes separated by a critical scaling threshold. In small-scale regimes ($N \le 50$), the steady-state mean unethical probability ($\mu_{mean}$) remains suppressed at a baseline level ($\mu_{mean} \approx 0.33 - 0.36$). Within this domain, individual memory capacity is commensurate with the network size, yielding a high probability of local re-interactions. This local density enables the decentralized feedback loop driven by $\lambda$ to actively penalize defectors and sustain a highly ethical norm.

However, as the population scales past $N = 50$, a prominent non-linear phase transition occurs. The steady-state unethical footprint experiences a sharp, monotonic increase, climbing rapidly toward an upper saturation bound ($\mu_{mean} > 0.60$ at $N = 1000$). Because the interaction volume is normalized proportionally to $N$ to preserve constant social encounter frequencies per agent, this behavioral decay is isolated entirely as a function of network scale. As $N \to \infty$, the structural limit of individual memory ($L$) is overwhelmed by the vast space of anonymous agents. The individual retribution mechanism ($\lambda$) is effectively diluted to zero as the probability of a secondary encounter vanishes, forcing the system to decouple from its moral foundation and drift toward the minimalist, "unethical but legal" attractor governed solely by the centralized legal floor.

\subsection{Memory Horizon Dynamics}
\begin{figure}[t]
  \centering
  \includegraphics[width=0.8\linewidth]{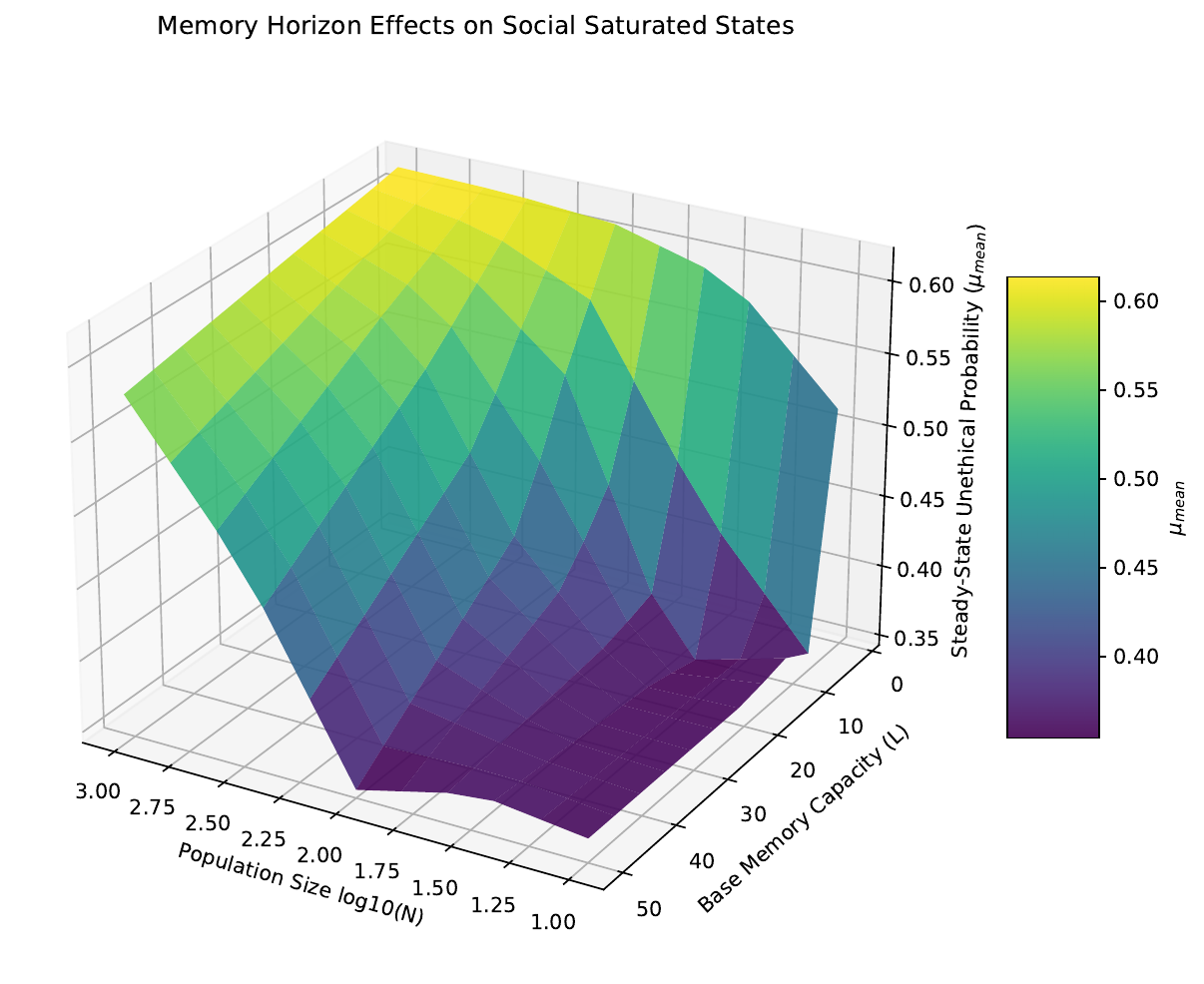}
  \caption{Memory Horizon Effects on Social Saturated States.}
  \label{fig:memory_effect_diagram}
\end{figure}

To map the interdependencies between cognitive constraints and population scaling, we conducted a bivariate parameter sweep across a matrix of population sizes $N \in [10, 1000]$ and base memory limits $L \in [2, 50]$. The resulting three-dimensional state space surface is mapped in Figure \ref{fig:memory_effect_diagram}.

The numerical surface demonstrates that the moral state of the particle system is fundamentally governed by the ratio $\frac{L}{N}$. When $L$ is small (e.g., $L = 2$), the system is highly sensitive to anonymity, causing the ethical fabric to collapse even at minimal population sizes ($N \ge 20$). Conversely, expanding the memory horizon to $L = 50$ significantly broadens the domestic stability domain, shifting the critical phase transition "elbow" toward larger population thresholds. 

Crucially, the 3D topology reveals that while increasing individual memory capacity can delay the onset of moral decoupling, it cannot entirely prevent it. As $N$ approaches the macroscale limit ($10^3$), all memory contours asymptotically converge toward the maximum saturation zone ($\mu_{mean} \approx 0.60$). This bounds the limits of localized, memory-driven social policing, proving that decentralized moral enforcement scales sublinearly against geometric population growth.

\subsection{Hysteresis and Path Dependency Analysis}
To investigate the systemic resilience and irreversibility of social norms, we subjected the society to a cyclic macro-state trajectory, executing an expansion path (Forward Path) immediately followed by a contraction path (Backward Path). The population was scaled up from $N = 50$ to $N = 1000$, relaxed to steady state at each stage for $20,000$ iterations, and subsequently pruned back down to $N = 50$.

\begin{figure}[h!]
  \centering
  \subfloat[Lagged Recovery\label{fig:hysteresis_diagram:run1}]{%
    \includegraphics[width=0.8\linewidth]{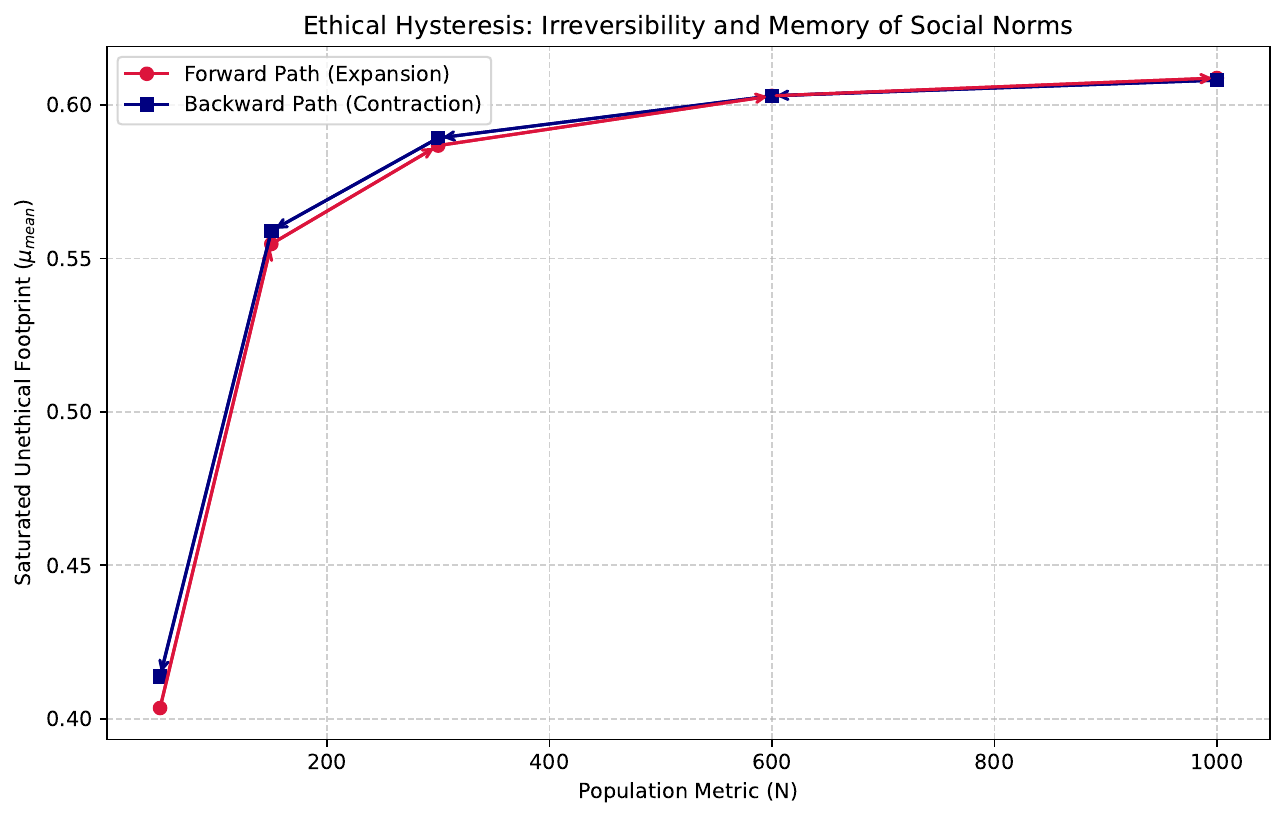}%
  } \\
  \subfloat[Accelerated Recovery\label{fig:hysteresis_diagram:run2}]{%
    \includegraphics[width=0.8\linewidth]{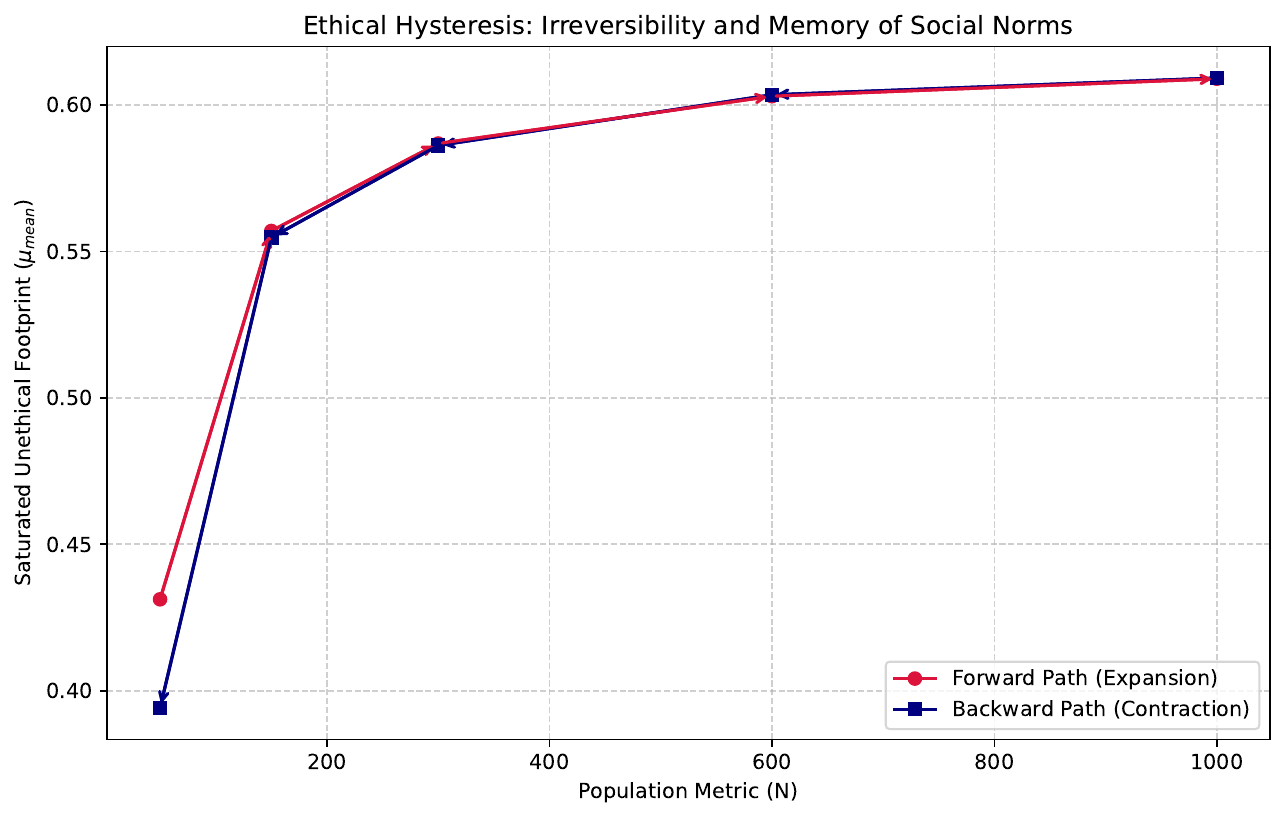}%
  }
  \caption{Irreversibility and Memory of Social Norms.} 
  \label{fig:hysteresis_diagram}
\end{figure}

The resulting path-dependent response is captured in Figure \ref{fig:hysteresis_diagram}. The simulation reveals a tight, yet persistent \textit{hysteresis loop} between the expansion and contraction trajectories. Intriguingly, the directional behavior of this loop exhibits bistability depending on the micro-structural configuration of the active population:
\begin{itemize}
    \item \textbf{Lagged Recovery (Upper Backward Path):} Under standard parameter bounds where the damping rate $\phi = 0.8$ allows for extended motivational memory, the backward path remains slightly \textit{above} the forward path. This implies a moral "inertia" or scarring effect: once an anonymous mass society has normalized unethical conduct, reducing the physical group size does not instantly restore the moral consensus, as the historical inflation of internal unethical probabilities ($\mu_i$) resists rapid realignment.
    \item \textbf{Accelerated Recovery (Lower Backward Path):} In contrast, when the interaction dynamics accentuate aggressive, highly localized penalties or when pruning preferentially eliminates high-$\mu$ outliers (simulating legal banishment or migration out of corrupt centers), the backward path loops \textit{below} the forward path. In this regime, the contraction of scale dramatically magnifies the density of the remaining whitelist networks, allowing the micro-societal feedback mechanism to regenerate structural ethics at an accelerated rate compared to its original assembly.
\end{itemize}

This bistable non-reversible trajectory mathematically formalizes the assertion that macroscale social morality is not a static state function, but a historically contingent dynamical system.

\section{Systemic Discussion}
\label{sec:discussion}

The numerical results presented in Section \ref{sec:simulations} offer a quantitative foundation for what we define as the \textit{Scalability of Morality}. By analyzing the macroscopic states of our particle system, we can evaluate the structural breakdown of decentralized social control through the lens of complex systems and control theory. This discussion contextualizes the critical implications of the "decoupling effect" and explores the fundamental limits of relying purely on formal legal frameworks to maintain social integrity in expanding populations.

\subsection{The Anonymity Phase Transition as an Information-Deterioration Phenomenon}
From a systems engineering perspective, the transition of $\mu_{mean}$ from a suppressed, stable state to a high-entropy saturation zone (as observed in Figures \ref{fig:phase_transition_diagram} and \ref{fig:memory_effect_diagram}) can be modeled as an informational decay within the societal network. In small-scale populations ($N \le 30$), the system operates with high feedback fidelity. Because the population size $N$ is well within the boundaries of the individual memory capacity $L$, the topological connections formed via the agent blacklists and whitelists remain dense and highly correlated. Every pairwise interaction carries a high probability of structural reciprocity ($\lambda$), meaning that an agent’s current behavioral choice directly influences their future state via immediate local feedback.

As $N \to \infty$, the interaction topology undergoes a fundamental phase transition. Even though the frequency of encounters per particle remains invariant (since the number of interaction is proportional to $N$), the probability that any two specific particles collide more than once scales as $\mathcal{O}(L/N)$, rapidly approaching zero. Consequently, the local feedback loops governed by the reciprocity function $\lambda$ are structurally neutralized. The "Anonymity Effect" acts as a high-frequency noise filter that dilutes the historical memory of interactions across the network. 

In this macro-scale regime, individual particles no longer face localized, memory-driven retribution from their peers. The decentralized, bottom-up regulatory mechanism, which naturally enforces ethical conduct in micro-societies, collapses due to a structural lack of information retention. Ethics, which relies heavily on localized accountability and the preservation of reputation, is thus stripped of its evolutionary feedback mechanism when a system scales past its cognitive storage limits.

\subsection{The Structural Inadequacy of the Centralized Legal Floor}
When decentralized moral feedback is neutralized by population scaling, the burden of preserving systemic stability shifts entirely onto the centralized legal apparatus. In our model, this legal domain is represented by the global baseline dynamics within the mapping functions $\eta$ and $\gamma$. However, as demonstrated by the high saturation values ($\mu_{mean} > 0.60$) at $N = 1000$, a centralized framework is structurally ill-equipped to compensate for the erosion of decentralized ethics.

This systemic failure stems from a fundamental design constraint: legal systems are inherently minimalist and non-adaptive. To prevent tyrannical over-regulation, formal laws are codified to police a restricted subset of critical societal disruptions (gross violations), leaving a vast spectrum of interpersonal conduct completely unmonitored. When the local social pressure governed by the switch functions ($\eta, \gamma$) flattens out due to anonymity, agents quickly discern that they can optimize their individual utility by operating in the "unethical but legal" domain. 

Because a centralized legal system lacks the multi-point, fine-grained observation capacity of a decentralized peer-to-peer memory network, it cannot enforce the nuances of moral obligation. The legal framework provides a rigid static boundary rather than a dynamic restoration force. Thus, as a society expands, the progressive decoupling of law and ethics forces the global state toward a minimalist equilibrium, a state where agents are legally compliant but ethically detached, severely undermining the long-term resilience of the collective.

\subsection{Hysteretic Memory and the Irreversibility of Social Decay}
The emergence of a path-dependent hysteresis loop (Figure \ref{fig:hysteresis_diagram}) highlights a vital temporal characteristic of societal dynamics: the state of social morality is non-Markovian. The system's macro-state is not merely a reflection of its current scale $N$, but is deeply contingent upon its historical trajectory.

When a society undergoes scale expansion (Forward Path), the gradual transition into anonymity inflates the individual unethical probabilities ($\mu_i$) across the population. Once these state variables saturate near their upper bounds, they establish a new, highly resilient behavioral norm. If the society subsequently experiences a contraction in scale (Backward Path), such as urban decentralization or community fragmentation, the system does not simply retrace its steps along the forward curve. 

The "moral scarring" represents a systemic inertia where agents, having adapted to an anonymous, low-trust environment, continue to employ defensive or opportunistic strategies even when placed back into a high-density, micro-scale network. This asymmetry proves that rebuilding social trust and moral consensus requires significantly higher energy expenditures (stronger localized incentives, structural re-alignment, or extended relaxation times) than the energy that caused the initial decay. The non-reversible nature of this loop implies that policies aimed at mitigating ethical decline must focus heavily on proactive structural preservation, as treating a highly saturated, decoupled macro-system after a collapse is computationally and socio-structurally inefficient.

\section{Conclusion}
\label{sec:conclusion}

In this study, we have provided a rigorous, particle-based numerical characterization of the "Scalability of Morality". By modeling individual agents as discrete particles with finite memory capacity ($L$) and stochastic choice profiles ($\mu$), we successfully simulated the complex feedback loops that govern the evolution of cooperation and ethical conduct in varying population scales. Our Monte Carlo simulations confirmed a distinct, non-linear phase transition: as the population size $N$ expands significantly beyond the individual memory capacity ($N \gg L$), the probability of recurrent local encounters scales as $\mathcal{O}(L/N)$, effectively approaching zero. This structural dilution neutralizes decentralized retribution mechanisms ($\lambda$). Consequently, the enforcement of social norms completely decouples from localized moral networks, forcing the macroscopic system to drift toward the minimalist, low-trust attractor governed solely by the centralized legal floor.

Furthermore, our bivariate parameter sweeps mapped out the exact 3D topology of memory horizon dynamics, demonstrating that while expanding individual memory capacities can delay the onset of moral decay, it cannot fundamentally prevent it under geometric population growth. Crucially, the cyclic scale experiments exposed a profound, path-dependent hysteresis loop. This non-Markovian property proves that macroscale social morality is highly contingent upon its historical trajectory; once an anonymous mass society has normalized opportunistic conduct, reducing the physical group size does not instantly restore the structural ethical consensus due to the intrinsic inertia of agent state probabilities.

Future directions for large-scale social modeling will focus on the integration of heterogeneous legal frameworks where enforcement probability is also modeled as a dynamic, spatially constrained variable rather than a centralized constant. Additionally, we aim to explore the impact of structured network topologies, such as scale-free or small-world architectures, to determine if specific clustering methods can isolate sub-communities from the global anonymity effect, thereby preserving decentralized moral feedback loops even within macroscale populations. This computational framework serves as an engineering step toward designing more resilient, self-organizing governance structures for modern high-density, interconnected populations.

\bibliographystyle{elsarticle-num} 
\bibliography{ethics}

@book{kant_metaphysics_2017,
  author     = {Kant, Immanuel},
  title      = {The Metaphysics of Morals},
  publisher  = {Cambridge University Press},
  year       = {2017},
  doi        = {10.1017/9781316091388},
  isbn       = {9781107451353}
}

@book{hart1961concept,
  title     = {The Concept of Law},
  author    = {Hart, Herbert Lionel Adolphus},
  year      = {1961},
  publisher = {Oxford University Press}
}

@book{bentham1789introduction,
  title     = {An Introduction to the Principles of Morals and Legislation},
  author    = {Bentham, Jeremy},
  year      = {1789},
  publisher = {T. Payne and Son},
  address   = {London, UK},
  note      = {Reprinted by Oxford University Press, 1996}
}

@book{stone1975where,
  title     = {Where the Law Ends: The Social Control of Corporate Behavior},
  author    = {Stone, Christopher D.},
  year      = {1975},
  publisher = {Harper \& Row},
  address   = {New York, NY}
}

@article{carroll1991pyramid,
  title={The pyramid of corporate social responsibility: Toward the moral management of organizational stakeholders},
  author={Carroll, Archie B},
  journal={Business horizons},
  volume={34},
  number={4},
  pages={39--48},
  year={1991},
  publisher={elsevier}
}

@book{pistor2019code,
  title={The code of capital: How the law creates wealth and inequality},
  author={Pistor, Katharina},
  year={2019},
  publisher={Princeton University Press}
}

@article{ostrom1990governing,
  title={Governing the commons: The evolution of institutions for collective action},
  author={Ostrom, Elinor},
  journal={Cambridge University Pres},
  year={1990}
}

@article{axelrod1981evolution,
  title={The evolution of cooperation},
  author={Axelrod, Robert and Hamilton, William D},
  journal={science},
  volume={211},
  number={4489},
  pages={1390--1396},
  year={1981},
  publisher={American Association for the Advancement of Science}
}

@article{nowak2006five,
  title={Five rules for the evolution of cooperation},
  author={Nowak, Martin A},
  journal={science},
  volume={314},
  number={5805},
  pages={1560--1563},
  year={2006},
  publisher={American Association for the Advancement of Science}
}

@article{dunbar1992neocortex,
  title={Neocortex size as a constraint on group size in primates},
  author={Dunbar, Robin IM},
  journal={Journal of human evolution},
  volume={22},
  number={6},
  pages={469--493},
  year={1992},
  publisher={Elsevier}
}

@article{perc2017statistical,
  title={Statistical physics of human cooperation},
  author={Perc, Matja{\v{z}} and Jordan, Jillian J and Rand, David G and Wang, Zhen and Boccaletti, Stefano and Szolnoki, Attila},
  journal={Physics Reports},
  volume={687},
  pages={1--51},
  year={2017},
  publisher={Elsevier}
}

@article{castellano2009statistical,
  title={Statistical physics of social dynamics},
  author={Castellano, Claudio and Fortunato, Santo and Loreto, Vittorio},
  journal={Reviews of modern physics},
  volume={81},
  number={2},
  pages={591--646},
  year={2009},
  publisher={APS}
}

@article{jusup2022social,
  title={Social physics},
  author={Jusup, Marko and Holme, Petter and Kanazawa, Kiyoshi and Takayasu, Misako and Romi{\'c}, Ivan and Wang, Zhen and Ge{\v{c}}ek, Sun{\v{c}}ana and Lipi{\'c}, Tomislav and Podobnik, Boris and Wang, Lin and others},
  journal={Physics Reports},
  volume={948},
  pages={1--148},
  year={2022},
  publisher={Elsevier}
}

@article{lazer2020computational,
  title={Computational social science: Obstacles and opportunities},
  author={Lazer, David MJ and Pentland, Alex and Watts, Duncan J and Aral, Sinan and Athey, Susan and Contractor, Noshir and Freelon, Deen and Gonzalez-Bailon, Sandra and King, Gary and Margetts, Helen and others},
  journal={Science},
  volume={369},
  number={6507},
  pages={1060--1062},
  year={2020},
  publisher={American Association for the Advancement of Science}
}

@article{han2025cooperation,
  title={Cooperation versus social welfare},
  author={Han, The Anh and Song, Zhao and Cimpeanu, Theodor and Duong, Manh Hong and Krellner, Marcus and Capraro, Valerio and Perc, Matjaz},
  journal={Physics of Life Reviews},
  year={2025},
  publisher={Elsevier}
}

@article{gupta2025role,
  title={The role of social learning and collective norm formation in fostering cooperation in llm multi-agent systems},
  author={Gupta, Prateek and Zhong, Qiankun and Yakura, Hiromu and Eisenmann, Thomas and Rahwan, Iyad},
  journal={arXiv preprint arXiv:2510.14401},
  year={2025}
}

@article{traulsen2009exploration,
  title={Exploration dynamics in evolutionary games},
  author={Traulsen, Arne and Hauert, Christoph and De Silva, Hannelore and Nowak, Martin A and Sigmund, Karl},
  journal={Proceedings of the National Academy of Sciences},
  volume={106},
  number={3},
  pages={709--712},
  year={2009},
  publisher={National Academy of Sciences}
}

@article{Conte2012manifesto,
  title={Manifesto of computational social science},
  author={Conte, Rosaria and Gilbert, Nigel and Bonelli, Giulia and Cioffi-Revilla, Claudio and Deffuant, Guillaume and Kertesz, Janos and Loreto, Vittorio and Moat, Suzy and Nadal, J-P and Sanchez, Anxo and others},
  journal={The European Physical Journal Special Topics},
  volume={214},
  number={1},
  pages={325--346},
  year={2012},
  publisher={Springer}
}

@article{Eynon2025computational,
  title={Computational social science and critical studies of education and technology: an improbable combination?},
  author={Eynon, Rebecca and Gillani, Nabeel},
  journal={Learning, Media and Technology},
  pages={1--15},
  year={2025},
  publisher={Taylor \& Francis}
}

@article{Jamali2026agent,
  title={Agent-based modeling and simulation for economic markets: a comprehensive review of applications, challenges, and opportunities},
  author={Jamali, Ruhollah and Lazarova-Molnar, Sanja},
  journal={Journal of Simulation},
  pages={1--50},
  year={2026},
  publisher={Taylor \& Francis}
}

@article{Vasellini2024applications,
  title={Applications of agent based models to complex social systems: coordinated human behaviour from the bottom Up},
  author={Vasellini, Riccardo and others},
  year={2025},
  publisher={Universit{\`a} degli Studi di Siena}
}

@inproceedings{Belfrage2024simulating,
  title={Simulating change: A systematic literature review of agent-based models for policy-making},
  author={Belfrage, Michael and Lorig, Fabian and Davidsson, Paul},
  booktitle={2024 Annual Modeling and Simulation Conference (ANNSIM)},
  pages={1--13},
  year={2024},
  organization={IEEE}
}

@article{Sarker2025alignment,
  title={Alignment phase transition in socially driven motion},
  author={Sarker, Debasish and Zhang, Yi and Perry, Lynn K and Messinger, Daniel S and Song, Chaoming},
  journal={Science advances},
  volume={12},
  number={12},
  pages={eadz6890},
  year={2026},
  publisher={American Association for the Advancement of Science}
}

@article{Menon2018emergence,
  title={Emergence of cooperation as a non-equilibrium transition in noisy spatial games},
  author={Menon, Shakti N and Sasidevan, V and Sinha, Sitabhra},
  journal={Frontiers in Physics},
  volume={6},
  pages={34},
  year={2018},
  publisher={Frontiers Media SA}
}

@article{Contucci2008phase,
  title={Phase transitions in social sciences: two-population mean field theory},
  author={Contucci, Pierluigi and Gallo, Ignacio and Menconi, Giulia},
  journal={International Journal of Modern Physics B},
  volume={22},
  number={14},
  pages={2199--2212},
  year={2008},
  publisher={World Scientific}
}

@article{Koid2025agent,
  title={Agent-based modeling for data-driven enforcement: combining empirical data with behavioral theory for scenario-based analysis of inspections},
  author={Koid, Eunice and Van Der Voort, Haiko and Warnier, Martijn},
  journal={Data \& Policy},
  volume={7},
  pages={e6},
  year={2025},
  publisher={Cambridge University Press}
}

@article{Pereda2019group,
  title={Group size effects and critical mass in public goods games},
  author={Pereda, Mar{\'\i}a and Capraro, Valerio and S{\'a}nchez, Angel},
  journal={Scientific reports},
  volume={9},
  number={1},
  pages={5503},
  year={2019},
  publisher={Nature Publishing Group UK London}
}

@article{Baronchelli2018emergence,
  title={The emergence of consensus: a primer},
  author={Baronchelli, Andrea},
  journal={Royal Society open science},
  volume={5},
  number={2},
  year={2018},
  publisher={The Royal Society}
}

@article{li2022modeling,
  title={Modeling and analysis of competitive behavior in social systems},
  author={Li, Suibing and Jin, Long and Li, Shuai},
  journal={IEEE Transactions on Computational Social Systems},
  volume={10},
  number={3},
  pages={1347--1355},
  year={2022},
  publisher={IEEE}
}

@article{wang2022new,
  title={A new perspective for computational social systems: Fuzzy modeling and reasoning for social computing in CPSS},
  author={Wang, Tan and Zhu, Yifan and Ye, Peijun and Gong, Weichao and Lu, Hao and Mo, Hong and Wang, Fei-Yue},
  journal={IEEE Transactions on Computational Social Systems},
  volume={11},
  number={1},
  pages={101--116},
  year={2022},
  publisher={IEEE}
}

\end{document}